\documentclass[12pt]{article}
\usepackage{epsfig}

\def\beq{\begin{equation}}
\def\eeq#1{\label{#1}\end{equation}}
\def\eeqn{\end{equation}}

\def\beqa{\begin{eqnarray}}
\def\eeqa#1{\label{#1}\end{eqnarray}}
\def\eeqan{\end{eqnarray}}

\let\bar=\overbar

\def\Dslash{\not{\hbox{\kern-4pt $D$}}}
\def\dslash{\not{\hbox{\kern-2pt $\del$}}}

\def\msb{{\bar{\ssstyle M \kern -1pt S}}}

\usepackage{fancyhdr,graphicx}
\def\Title#1{\begin{center} {\Large {\bf #1} } \end{center}}

\begin{document}

\Title{Effects of rotochemical heating on the thermal evolution of superfluid neutron stars }

\bigskip\bigskip


\begin{raggedright}

{\it Chun-Mei Pi\index{Med, I.}\\
Institute of Astrophysics\\
Huazhong Normal University\\
Wunhan 430079\\
P. R. China\\
{\tt Email: pcm1221@phy.ccnu.edu.cn}}
\bigskip\bigskip
\end{raggedright}

\section{Introduction}

With the development of observation technology, we have more opportunity to observe and analyze the thermal radiation from the compact star surfaces. The comparison between the predictions of theoretical study of the thermal evolution  and observations of thermal emission from the compact stars gives a potentially powerful method to study fundamental properties of superdense matter in compact star interiors.
\cite{hpy07, lp07}.

It is generally believed that dense matter can be superfluid at such high density and low temperature
\cite{ls01}. Baryon pairing (or quark pairing) strongly suppresses not only the neutrino emission but also the
heat capacity of nuclear matter. Meanwhile, Cooper pair formation and breaking also will effect the neutron star
cooling. Furthermore, as a neutron star spins down, some heating mechanisms may be present and will play important roles in the
thermal evolution of NS. They have been extensively discussed, for example, rotochemical heating
\cite{rei95,fr05}, mutual friction between superfluid and normal components of the star \cite{SL89}, crust
cracking \cite{cczc92} and the dissipation of rotational energy due to viscous damping \cite{zheng2006}. We
will focus on rotochemical heating in the following study.

It's well-known \cite{ykgh01} that for nonsuperfluid NS with the core composed of standard nuclear matter, we
have two distinct cooling regimes, slow and fast cooling. The transition from slow to fast cooling with
increasing M occurs in a very narrow mass range and is sharp. After considering the superfluidity, the Urca
neutrino emission and the NS heat capacity will be strongly suppresses, and neutron emission due to Cooper
pairing will occur \cite{frs76}. Then these effects will make NS warmer. Recently, when including nucleon
superfluidity and using APR EOS , \cite{gkyg05} found a particular scenario of neutron star cooling that the
theoretical cooling models of isolated middle-aged neutron stars can be divided into three distinct types: slow,
moderate and fast cooling. They neglected the heating effects, while we will take this effects into account
which may make the cooling scenario different.

\section{Nucleon Superfluidity}

Many microscopic studies of dense nucleon matter predict that below certain critical temperature nucleons will
be in superfluidity, and the critical temperatures, $T_{cp}$ and $T_{cn}$, depend sensitively on the model of
nucleon-nucleon interaction and many-body theory employed(see, e.g., \cite{yls99},\cite{yp04},for
references). It is widely accepted that there are three types nucleon superfluidities: singlet-state$^1$S$_0$
pairing of neutrons ($T_{\rm c}=T_{\rm cns}$) in the inner crust and outermost core; $^1$S$_0$ proton pairing
($T_{\rm c}=T_{\rm cp}$) in the core; and triplet-state ($^3$P$_2$) neutron pairing ($T_{\rm c}=T_{\rm cnt}$) in
the core.

Almost all contemporary theories predict some common features of nucleon superfluidity: $T_{\rm cn}(\rho)$ for
the singlet--state neutron SF has maximum at subnuclear densities in the crust and vanishes at $\rho \sim 2
\times 10^{14}$ g cm$^{-3}$ while $T_{\rm cn}(\rho)$ for the triplet--state neutron SF grows up at subnuclear
density, reaches maximum at $\rho=(2-3) \, \rho_0$ ($\rho_0=2.8 \times 10^{14}$ g cm$^{-3}$ is the saturation
density of nuclear matter) and decreases with $\rho$,  vanishing at $\rho \sim (3-5) \times 10^{15}$ g
cm$^{-3}$. $T_{\rm cp}$ also has maximum at several $\rho_0$ and vanishes at higher $\rho$ \cite{khy01,ykgh01}.
To describe the density dependences of $T_{\rm cn}$ and $T_{\rm cp}$, several authors propose phenomenological
models \cite{khy01,kyg02}.

The nucleon superfluidity suppresses the heat capacity and neutrino processes involving superfluid nucleon, and
what's more, superfluidity initiates an additional neutrino emission associated with Cooper pairing of nucleons
\cite{frs76}. Our codes of NS thermal evolution include all these effects.

\section{Rotochemical heating}

As neutron stars spin sown and contract, their structure and the chemical equilibrium state of beta processes change with the increasing density. Non-equilibrium reactions tend to restore equilibrium. During the relaxation to a new equilibrium state, the energy is stored. In addition, the chemical imbalance modifies the reaction rates \cite{haensel92}. If the departure from equilibrium is large enough, the net effect of the reactions is to increase the thermal energy at the expense of the stored chemical energy \cite{rg92}.

For $npe$ matter in NS, their relative concentrations are adjusted by direct Urca reactions,
\begin{eqnarray}
n \rightarrow p + e + \overline{\nu} , \\
p + e \rightarrow n  + \nu ,
\end{eqnarray}
and modified Urca reactions,
\begin{eqnarray}
n + N \rightarrow p + N + e + \overline{\nu} , \\
p + N + e \rightarrow n + N + \nu .
\end{eqnarray}
The departure from the chemical equilibrium can be quantified by the chemical imbalance:
\begin{equation}
\eta_{npe} = \delta\mu_n - \delta\mu_p - \delta\mu_e, \\
\end{equation}
where $\delta\mu_i = \mu_i - \mu_i^{eq}$ is the deviation from the equilibrium chemical potential of species $i$
at given pressure. In non-equilibrium state, neutrino emissivities and net reaction rates per unit volume of
Urca processes will be modified, which can be written as \cite{haensel92}
\begin{eqnarray}
Q_{\alpha}(n,T,\eta_\alpha) = Q_{\alpha}^{eq}(n,T)F_\ast(\frac{\eta_\alpha}{kT}) , \\
\Delta\Gamma_\alpha(n,T,\eta_\alpha) = \frac{1}{kT}Q_{\alpha}^{eq}(n,T)H_\ast(\frac{\eta_\alpha}{kT}) ,
\end{eqnarray}
where $Q_{\alpha}^{eq}$ is the neutrino emissivity in equilibrium and $\eta_\alpha$ is the chemical imbalance
due to reaction $\alpha$, $T$ is the local temperature, $n$ is the baryon number density, $k$ is Boltzmanm's
constant, and the expressions of $F_\ast$ and $H_\ast$ are given in the appendix of \cite{rei95}. Here we have
to announce that if we consider the property of nucleon superfluidity $Q_{\alpha}^{eq}$ includes the effects of
nucleon superfluidity by multiplying the reduction factors from \cite{ykgh01}. The total energy dissipation
rate per unit volume is
\begin{equation}
Q_H = \sum_\alpha\Delta\Gamma_\alpha\eta_\alpha .
\end{equation}

\section{Thermal Evolution}
\label{Extmincool}

Thermal evolution of NS can be distinguished three main stages:(i) the internal relaxation stage ($t \approx10$--100 yr; \cite{lvp94}), (ii) the neutrino stage (the neutrino luminosity $L_\nu \gg
L_\gamma$, $t \approx 10^5$ yr), and (iii) the photon stage ($L_\nu \ll L_\gamma$, $t > 10^5$ yr). After the
thermal relaxation, the redshifted temperature $T^\infty= T e^\phi$ becomes constant throughout the stellar
interior. The equation of thermal evolution can be written as
\begin{equation}
C_{V}\frac{dT_\infty}{dt} = -L_{\nu}^\infty -L_{\gamma}^\infty +L_H^\infty
\end{equation}
where $C_{V}$ is the total stellar heat capacity, $L_H^\infty$ is the total power released by the heating
mechanism, $L_{\nu}^\infty$ is the total power emitted as neutrinos, $L_{\gamma}^\infty$ is the power released
as thermal photon. These quantities are calculated as
\begin{eqnarray}
L_H^\infty &=& \int dV Q_H e^{2\phi},\\
L_{\nu}^\infty &=& \int dV Q_{\nu} e^{2\phi},\\
L_{\gamma}^\infty &=& 4\pi R^2 T_s^4 e^{2\phi_s}=4\pi R^2_\infty
(T_s^\infty)^4,
\end{eqnarray}
respectively, where $dV=4\pi r^2(g_{rr}^{1/2}) dr$ is the proper
volume element, $Q_{\nu}$ is the total neutrino emissivity
contributed by reactions. $\sigma$ is the Stefan-Boltzman constant,
R is the stellar coordinate radius, $\phi_{s}=\phi(R)$,
$R_\infty=Re^{-\phi_s}$ is the effective radius as measured from
infinity, and $T_s^\infty$ is the redshifted effective temperature.

When considering the rotochemical heating effect, we can write the time evolution of the temperature and the
chemical imbalances for NS constitute of $npe$ matter as:
\begin{equation}
\dot{T}_\infty = [M_D(\eta_{npe})L_D^\infty + M_M(\eta_{npe})L_M^\infty- L_{\nu}^{'}-L_{\gamma}^\infty]/C_{V}^\infty,\\
\end{equation}
\begin{equation}
\dot{\eta}^\infty_{npe}= -\frac{Z_{npe}}{kT}[L_D^\infty H_{D}(\eta^\infty_{npe},T)+L_M^\infty
H_{M}(\eta^\infty_{npe},T)] + 2W_{npe}\Omega\dot{\Omega},
\end{equation}
where $L_{\nu}^{'} = L^\infty_{NN} + L^\infty_{co}$, $L^\infty_{NN}$, $L^\infty_{co}$, $L_D^\infty$ and
$L_M^\infty$ are the neutrino luminosities of neutrino emission processes including in our thermal evolution
codes (neutrino bremsstrahlung in nucleon-nucleon scattering, neutrino emission due to Cooper pairing of
superfluid nucleons, direct and modified Urca reactions) respectively, the functions $M$ and $H$ quantify the
effect of reactions towards restoring chemical equilibrium, the subscripts D and M denote direct and modified
Urca reactions, the expression of $Z_{npe}$ relating to the structure of rotating NS is in \cite{fr05}, the
scalar $W_{npe}$ quantify the departure from equilibrium due to the change in the centrifugal force
($\propto\Omega\dot{\Omega}$) \cite{fr05,rfj07}.

\begin{figure}[htb]
\begin{center}
\epsfig{file=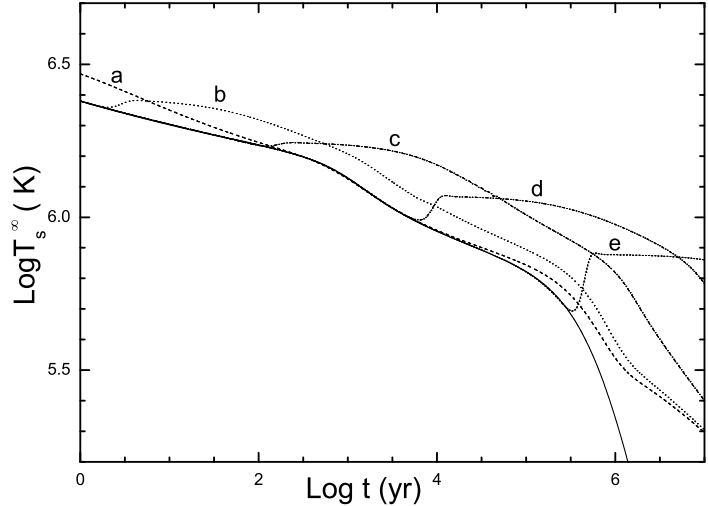,height=80mm}
\caption{Thermal evolution curves of $1.6 M_\odot$ superfluid NS for no heating
    (solid line) or rotochemical heating with magnetic field strengths (a) $B=10^{13}G$,
    (b) $B=10^{12}G$, (c) $B=10^{11}G$, (d) $B=10^{10}G$, (e) $B=10^{9}G$. The initial spin period is taken to be $1 ms$.}
\label{fig:1.6 different B}
\end{center}
\end{figure}
\begin{figure}[htb]
\begin{center}
\epsfig{file=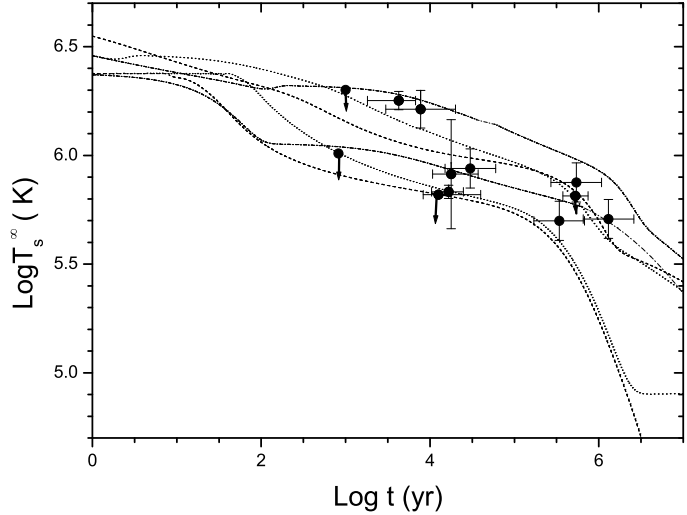,height=80mm}
\caption{Thermal evolution curves of $1.68 M_\odot$ and $1.7 M_\odot$ superfluid NS with rotochemical heating for different magnetic fields $B=10^{13}G$ (dash lines), $B=10^{12}G$ (dot lines),
$B=10^{11}G$ (dash-dot lines). The initial spin period is taken to be $1 ms$.}
\label{fig:2}
\end{center}
\end{figure}

We plot the thermal evolution curves of $1.6 M_\odot$ superfluid NS for no heating (solid line) or rotochemical
heating with various magnetic field strengths ($10^{9} - 10^{13} G$) in Figure 1. And the analogs figures can be
seen in \cite{rei95}. However, those in \cite{rei95} didn't include the effects of nucleon superfluidity. We
can see that rotochemical heating increases considerably the surface temperature of stellar whose interior
contains superfluid nucleon. It's quite clear that the thermal evolution curves are strongly depend on the
magnetic field strengths, which is related with the properties of MDR. It is obvious that the stronger the
magnetic field is the earier the interesting effects of rotochemical heating occur. For the strongest field
($B=10^{13}G$), the spin-down is rapid and the heating is distinct at early times while for the lowest field
($B=10^{9}G$) the heating has little effect until very late times. And the most interesting heating effects for
middle-aged stellars are those for intermediate range ($10^{10} - 10^{12} G$) in which the magnetic field of
observational sources lie.

The thermal evolution curves of $1.68 M_\odot$ and $1.7 M_\odot$ superfluid NSs with
rotochemical heating for different magnetic fields ($10^{11} - 10^{13} G$) are presented in Figure 3. We can easily see that our model is consistent with the observation data, and due to the different magnetic fields, the vacancy region in Figure 6 of \cite{gkyg05} will no longer exist which means three distinct cooling types proposed in \cite{gkyg05} become ambiguous.

\def\Discussion{
\setlength{\parskip}{0.3cm}\setlength{\parindent}{0.0cm}
     \bigskip\bigskip      {\Large {\bf Discussion}} \bigskip}
\def\speaker#1{{\bf #1:}\ }
\def\endDiscussion{}

\Discussion

We have studied the thermal evolution behaviors of rotating NSs with nucleon cores including the effects of
nucleons superfluidity and the rotochemical heating.

There are some shortcomings in our work. Firstly, we neglect processes and the heat capacity due to the lattice
of ions in the crust. Secondly, we ignore the effect of superfluidity on the reaction rates. Like the neutrino
emissivity, the reaction rates should be suppressed by superfluidity unless the departure from chemical
equilibrium $\eta$ exceeds the sum of the energy gap of the participating baryons\cite{rei97}.
\endDiscussion

\end{document}